\documentclass[aps,prb,superscriptaddress,floatfix,twocolumn]{revtex4-1}
\usepackage{amssymb}
\usepackage{amsmath}
\usepackage{wasysym}
\usepackage{verbatim}
\usepackage[colorlinks,bookmarks=false,citecolor=blue,linkcolor=red,urlcolor=blue]{hyperref}
\usepackage{times}
\usepackage{color}
\usepackage{graphicx}
\usepackage{epsfig}
\usepackage{float}
\usepackage{bbold}
\usepackage{tabularx}
\usepackage{natbib}

\pdfoutput=1

\newcommand{\Eref}[1]{Equation (\ref{#1})}

\newcommand{\Fref}[1]{Figure \ref{#1}}
\newcommand{\Sref}[1]{Section \ref{#1}}
\newcommand{\tr}{\mathrm{tr}}
\newcommand{\etal}{{\it et al.}~}
\newcommand{\up}{\uparrow}
\newcommand{\down}{\downarrow}
\newcommand{\kF}{k_\mathrm{F}}

\newcommand{\ident}{\mathbb{1}}
\begin{document}

\title{An entropy perspective on the thermal crossover in a fermionic Hubbard chain}

\author{Lars Bonnes}
\email{lars.bonnes@uibk.ac.at}
\affiliation{Institute for Theoretical Physics, University of Innsbruck, A-6020 Innsbruck, Austria.}

\author{Hannes Pichler}
\affiliation{Institute for Theoretical Physics, University of Innsbruck, A-6020 Innsbruck, Austria.}
\affiliation{Institute for Quantum Optics and Quantum Information of the Austrian Academy of Sciences, A-6020 Innsbruck, Austria.}

\author{Andreas M. L\"auchli}
\affiliation{Institute for Theoretical Physics, University of Innsbruck, A-6020 Innsbruck, Austria.}

\date{\today}

\begin{abstract}
We study the Renyi entropy in the finite temperature crossover regime of a Hubbard chain using quantum Monte Carlo.
The ground state entropy has characteristic features such as a logarithmic divergence with block size and $2\kF$ oscillations that are a hallmark of its Luttinger liquid nature.
The interplay between the (extensive) thermal entropy and the ground state features is studied and we analyze the temperature induced decay of the amplitude of the oscillations 
as well as the scaling of the purity. Furthermore, we show how the spin and charge velocities can be extracted from the temperature dependence of the Renyi entropy, bridging 
our findings to recent experimental proposals on how to implement the measurement of Renyi entropies in cold atom system.
Studying the Renyi mutual information, we also demonstrate how constraints such as particle number conservation can induce persistent correlations visible in the mutual information
even at high temperature.
\end{abstract}

\maketitle 


\section{Introduction}
Entanglement in quantum systems is a manifestation of inherent quantum correlations between different degrees of freedom~\cite{horodecki09}.
Over the last decades, quantum information theory lead to a deeper understanding of how to characterize and quantify entanglement in few and many body systems with a wide range of applications in condensed matter and cold atomic systems.
One of the most striking features for ground states for a wide class of physics Hamiltonians is the area law~\cite{eisert10a} for the entanglement entropy stating that the entanglement between two subsystems is determined by their \textit{boundary} -- alike the entropy of a black hole being proportional to the size of its horizon~\cite{bombelli86} -- and not the their volume.
This has striking consequences for gapped one dimensional quantum systems where the entanglement is independent of block size~\cite{hastings07} consequently resulting in the success of the density matrix renormalization group~\cite{white92,white93,peschel99,schollwoeck05} (DMRG) and its descendants.
Measuring entanglement entropies in real-life experiments, however, is quite challenging since the Renyi entropies, for instance, are non-linear functionals of the reduced density matrix.
The latter has actually been measured using state tomography in trapped ions~\cite{haffner05,blatt08} and superconducting qubits~\cite{steffen06} for few particles but scaling this approach to systems with possibly hundreds or thousands of degrees of freedoms is challenging~\cite{devoret13,monroe13}.

Cold atom systems are promising candidates to implement scalable entanglement measurements as they provide the possibility to engineer interacting quantum (lattice) systems and, as recent theoretical proposal show~\cite{ekert02,cardy11,abanin12,daley12,pichler13}, directly access the Renyi entropies of the system.
In particular, as put forward by Daley \etal~\cite{daley12}, the $n$th Renyi entropy for a bosonic lattice system can be inferred from the expectation value of the swap operator by performing correlated site-resolved density measurements on $n$ different realizations of the systems.
These measurements can be implemented by using a quantum gas microscope, as it has already been realized by different experimental groups~\cite{bakr09,sherson10}, that allows for a single-site resolved detection of local occupations.
In a recent paper~\cite{pichler13}, we generalized this scheme to the case of fermionic atoms where special care has to be taken of phase factors from the braiding of two identical fermions.
These new experimental tools can not only test theoretical concept such as dynamical entropy generation in quantum quenches~\cite{calabrese05} or the conformal structure of critical one-dimensional systems but also provide insight in outstanding theoretical questing as corrections to the area law in $(2+1)$ dimensions~\cite{fradkin06,ryu06,metlitski09,tagliacozzo09,kallin11,metlitski11,humeniuk12,ju12} or topologically ordered systems~\cite{kitaev06,levin06}.

Whereas conformal field theory is a powerful tool to obtain the entanglement properties of many-body systems in one dimension, the crossover from a highly entangled many-body wave function to a thermal state as temperature is increased is much less explored.
With the prospect of future experiments that will operate in this intermediate regime a deeper understanding of the interplay between entropy and thermal entropy is of great interest.

To address these questions, we use recent finite-temperature quantum Monte Carlo methods complemented by ground-state DMRG to access the Renyi entropies in a fermionic Hubbard chain.
We can track the ground-state features such as the logarithmic divergence of the entanglement entropy and the spatial $2\kF$ oscillations over a wide range of temperatures and demonstrate how the low energy Luttinger liquid can be probed beyond the determination of the central charge by extracting the spin and charge velocities from the temperature behavior Renyi entropies.
We also study the mutual information and show how constraints such as particle number conservation affect the fluctuations that lead to finite mutual information of a bipartite splitting even at infinite temperatures.
From the perspective of prospective experiments, these results provide valuable insight on the temperature regimes where effective ground state entanglement will be visible and what information about the true ground state can be inferred from thermal Renyi entropies. 

This paper is structured as follows.
In \Sref{sec:model} and \ref{sec:entr}, the one-dimensional Hubbard model and previous results for the ground-state entanglement entropy and thermodynamic properties of the underlying Luttinger liquid are introduced.
An additional measure for mixed states, namely the Renyi mutual information, is introduced in \Sref{sec:mutInf} followed by a short discussion of the QMC methods employed to extract the Renyi entropies for thermal states in \Sref{sec:numerics}.
The simulation results are presented in \Sref{sec:result} where we show the finite-temperature crossover in \Sref{sec:crossover} and the extraction of the Luttinger velocities in \Sref{sec:velocities}.
Finally, \Sref{sec:resMI} discusses the mutual information for constrained (canonical) as well as unconstrained system.

\section{Model and methods}
\subsection{One dimensional Hubbard model}
\label{sec:model}
We consider spinful fermions on a chain of length $L$ with open boundary conditions.
One of the simplest interacting models is the Hubbard model with nearest-neighbor hopping and on-site repulsive charge interaction, $U\geq 0$, reading
\begin{equation}
 H=-t \sum_{\sigma=\up,\down} \sum_{i=1}^{L-1} \left(c_{i,\sigma}^\dagger c_{i+1,\sigma} + \mathrm{h.c.} \right) 
+ U \sum_{i=1}^L n_{i,\up} n_{i,\down}.
\end{equation}
The fermionic operators $c^\dagger_{i,\sigma}$ and $c_{i,\sigma}$ create or annihilate a particle with spin $\sigma$ on site $i$ and $n_{i,\sigma}=0,1$ counts the number of spin-$\sigma$ fermions on that lattice position.
Furthermore, we will only consider the spin balanced sector, i.e. $n_\up = n_\down$ ($n_\sigma = 1/L \sum_i n_{i,\sigma}$).

The Hubbard model is exactly solvable via the Bethe ansatz~\cite{lieb68,essler05} and exhibits a metallic ground state for generic fillings $n=n_\up+n_\down \ne 1$ that is described by a two-channel Luttinger liquid of the spin and charge degrees of freedom with velocities $v_\sigma$ and $v_\rho$.~\cite{woynarovich89,frahm90,essler05}
At half filling, on the other hand, a charge gap opens for $U>0$ whereas the spin degrees of freedom remain critical -- the system forms a Mott insulator.
Although the charge degrees of freedom are frozen out the dynamics of the spins give rise to an effective antiferromagnetic superexchange coupling $J=4t^2/U$ of adjacent spins.

\subsection{Entropy}
\label{sec:entr}
The $n$th Renyi entropy of a subsystem $A$ containing $l_A$ lattice sites is defined in terms of the moments of the reduced density matrix $\rho_A = \tr_{\bar A} \rho$ as
\begin{equation}
 S_n(A) = \frac{1}{1-n} \log \tr \rho_A^n.
\label{eq:Renyi}
\end{equation}
Here $\bar A$ denotes the complement of $A$ and $\rho$ is a not necessarily pure density matrix.
For $n \rightarrow 1$ one recovers the von Neumann entropy $S_\mathrm{vN}=-\tr [\rho_A \log \rho_A]$.

Insight into the entanglement properties and the logarithmic corrections to the area law for ground states of the Hubbard chain is provided by conformal field theory.~\cite{calabrese04,calabrese09}
The leading contribution for a bipartite splitting of a chain with open boundary conditions reads~\cite{calabrese09,calabrese10, calabrese10a,fagotti11} 
\begin{equation}
S^\mathrm{CFT}_n(A) = \frac{c}{12} \left(1+\frac{1}{n} \right) \log 
2 l^{\prime}_A + \mathrm{const.},
\label{eq:RenyiCFT}
\end{equation}
where $l^{\prime}_A = L/\pi \sin(\pi l_A/L)$ is the chord distance of block $A$.
The prefactor of the logarithmic divergence is sensitive to the central charge and readily distinguishes between the the $c=1$ Mott insulator~\cite{woynarovich89} and the semi-direct product of the two Luttinger liquids governing the metallic phase resulting in an effective central charge of $c=2$~\cite{woynarovich89,frahm90}.
In the case of open boundary conditions considered here, non-leading lattice effects will lead to $2\kF$ oscillations even for the von Neumann entropy -- here $\kF$ is the Fermi momentum -- whose decay with $l'_A$ is governed by the Luttinger parameters~\cite{laflorencie06,calabrese10,xavier11,fagotti11}.
Furthermore, there exist other corrections to \Eref{eq:RenyiCFT} such as shell filling effects.~\cite{essler13}

Conformal field theory can also access the entropy of systems at finite temperature and the von Neumann entropy for the Hubbard chain in the thermodynamic limit is given as~\cite{calabrese04,korepin04}
\begin{multline}
 S_\mathrm{vN}(A) = \frac{1}{3} \log \left[ \frac{v_s}{\pi T} \sinh\left( \frac{\pi T l_A}{v_s}\right) \right]
\\+ \frac{1}{3} \log \left[ \frac{v_c}{\pi T} \sinh\left( \frac{\pi T l_A}{v_c}\right) \right],
\label{eq:vNHubbard}
\end{multline}
where $v_c$ and $v_s$ denote charge and spin velocities.
This formula interpolates between the logarithmic growth with block size for $T \rightarrow 0$ and the volume law scaling $S_\mathrm{vN} \propto l_A$ of the thermal entropy.
A general expression for the Renyi entropies for a system of finite size \textit{and} finite temperature has, however, only been derived for massless Dirac fermions~\cite{azeyanagi08}.
Also, Ref.~\onlinecite{song12} discusses the finite-temperature crossover in the finite-size XX-chain using correlation matrix methods.

To make a connection to the thermodynamics of the Hubbard model, we reexpress the thermal Renyi entropy for a mixed state $\rho = \exp(-H/T)/Z$ of the entire system in terms of the Helmholtz free energy $F = -T\log Z$, namely
\begin{equation}
 S^\mathrm{th}_n(T) = -n \frac{ F(T/n) - F(T)}{T(1-n)}.
\end{equation}
Focusing on the Mott insulating phase where the charge degrees of freedom are gapped out, the low-temperature specific heat for a $l_A$ site system is just $C_V=\pi T l_A/(3 v_s)$ where $v_s = J_1(\pi U/(8t))/J_0(\pi U/(8t))$ is the spin velocity~\cite{coll74,takahashi74,schulz94} and $J_n(x)$ denote the modified Bessel functions of first kind.
From these considerations one infers that the thermal $n=2$ Renyi entropy in the $T\rightarrow 0$ limit takes the form
\begin{equation}
 S^\mathrm{th}_2(T) = \pi T l_A/(4 v_s).
\label{eq:ThermalRenyi}
\end{equation}

\subsection{Mutual information}
\label{sec:mutInf}
The Renyi entropies for thermal states do not obey an area law due to the extensive contribution of the thermodynamic entropy and in general provide no information about the amount of information one subsystem has about the other.
Unlike entanglement entropy, thermal entropy is merely a measure of our lack of knowledge about the micro state than of actual correlations.
The mutual information $I_n$, defined in terms of Renyi entropies as
\begin{equation}
 I_n(A,B)=S_n(A)+S_n(B)-S_n(A \cup B),
\label{eq:mutInf}
\end{equation}
on the other hand is sensitive towards correlations between two blocks~\cite{wolf08} $A$ and $B$,  i.e. not only two-point correlation functions that are used characterize possible (quasi) long-range order but all information one subsystem has about the other.
For the bipartition of a pure state one simply recovers the Renyi entropy, i.e. $I_n(A,\bar A) = 2 S_n(A)$, that is constrained by the area law.
In addition to that, it has also been shown that the mutual information obeys an area law even for thermal states~\cite{srednicki93,amico08,eisert10a}.

The mutual information for thermal (classical) states has attracted a lot of interest recently with respect to phase transitions in classical~\cite{wilms11,lau13} and quantum spin models~\cite{melko10}.
Furthermore, the $I_n(A,\bar A)$ is closely related to the excess entropy used in classical complexity theory as a measure for spatial memory of an infinite chain~\cite{feldman98}.
Thus, being able to access $I_2$ in the experiment also opens new directions of study apart from aspect of (quantum) entanglement such as the loss of mutual information in a quantum quench.~\cite{laeuchli08}

\subsection{Numerical Methods}
\label{sec:numerics}
The Renyi entropy can be accessed in QMC by using the idea the the $n$th moment of the reduced density matrix can be written in terms of the ratio of two partition functions as $\tr \rho_A^n = Z_n(A)/Z^n$.~\cite{calabrese04}
Here, $Z=\tr \exp(-H/T)$ is the usual partition function and $Z_n(A)$ lives on the $n$-sheeted Riemann surface with cuts along subsystem $A$ (see e.g. Ref.~\onlinecite{calabrese04,calabrese05} for an illustration).
Several schemes have been devised and applied to measure the partition function ratio using classical Monte Carlo~\cite{caraglio08,iaconis12,alba13}, valence-bond projector Monte Carlo~\cite{hastings10}, stochastic series expansion algorithms~\cite{melko10,isakov10,humeniuk12} or path-integral Monte Carlo~\cite{humeniuk12}.
Here, we follow the ideas of Ref.~\onlinecite{humeniuk12} of using a global topology update $Z_n(A) \leftrightarrow Z^n$ that allows us to directly sample $Z_n(A)/Z^n$ by simply counting how long in terms of Monte Carlo time the system is in one or the other world line topology.
The QMC simulations of the Hubbard chain at fixed filling are performed using a sign-problem free directed loop algorithm in the stochastic series expansion (SSE) framework~\cite{sandvik99b,syljuasen02,alet05} and projecting the measurements into the subspace of fixed density.
To avoid sampling problems due to small matching probabilities of the world lines across the cuts of $Z_n$, we use the \textit{increment trick}~\cite{hastings10} that splits the calculation for a block of size $l_A$ into the calculation of smaller blocks.
This can be done by expanding the partition function ratio as
$Z_n(l_A)/Z^n = [Z_n(l_A)/Z_n(l_A - \delta l)] ... [Z_n(2\delta l)/Z_n(\delta l)] [Z_n(\delta l)/Z_n(0)]$ 
and identifying $Z_n(0)=Z^n$.
Each quotient can be simulated independently and $\delta l \sim 5$ -- similar to the results by Humeniuk and Roscilde~\cite{humeniuk12} -- is found to be a good tradeoff between the number of increments and simulation efficiency.

It is noteworthy that $S_n(A \cup B)$ can be readily obtained for arbitrary intervals $A$ and $B$ in QMC since placing the cut(s) in $Z_n$ is arbitrary whereas the extraction of $I_2(A,B)$ for arbitrary blocks in DMRG is, in general, more involved.

In addition to our finite-$T$ QMC simulations DMRG~\cite{white92,white93,peschel99,schollwoeck05} is used to access the ground state properties of the Hubbard chain.
The results shown are well converged using up to 3000 basis states for the largest systems of $L=64$ sites.

\section{Finite-temperature crossover regime}
\label{sec:result}
\subsection{Entropy profiles and the purity}
\label{sec:crossover}

\begin{figure}[t]
\includegraphics[width=\columnwidth]{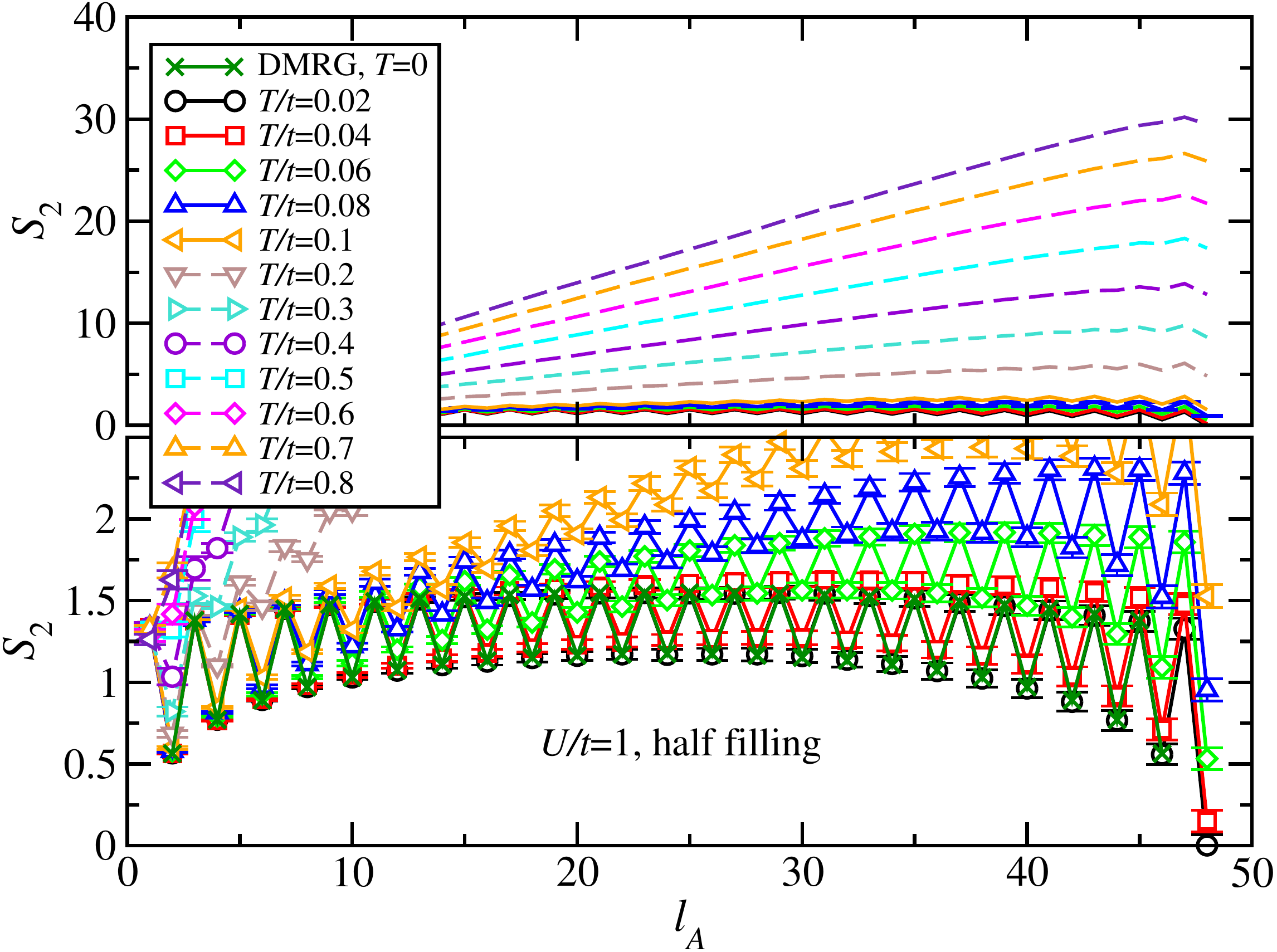}
\caption{(Color online) $n=2$ Renyi entropy profile for the Mott insulator at $U/t=1$ for a chain of length $L=48$.
}
\label{fig:EntropyProfileU1}
\end{figure}


Although the Luttinger liquid ground state of the Hubbard model is gapless in thermodynamic limit, a finite spatial extent and thus an infra-red cut-off leads to a gap to the first excited state that closes like $1/L$ for large system sizes~\cite{woynarovich87}.
Consequently, the ground state can effectively be probed even at finite temperatures as long as the temperature is well below the finite-size gap.
For larger temperatures, the low lying linearly dispersing excitations will be populated leading to a finite correlation length in the system as well as an extensive thermal contribution to the entropy.

\begin{figure}[t]
\includegraphics[width=\columnwidth]{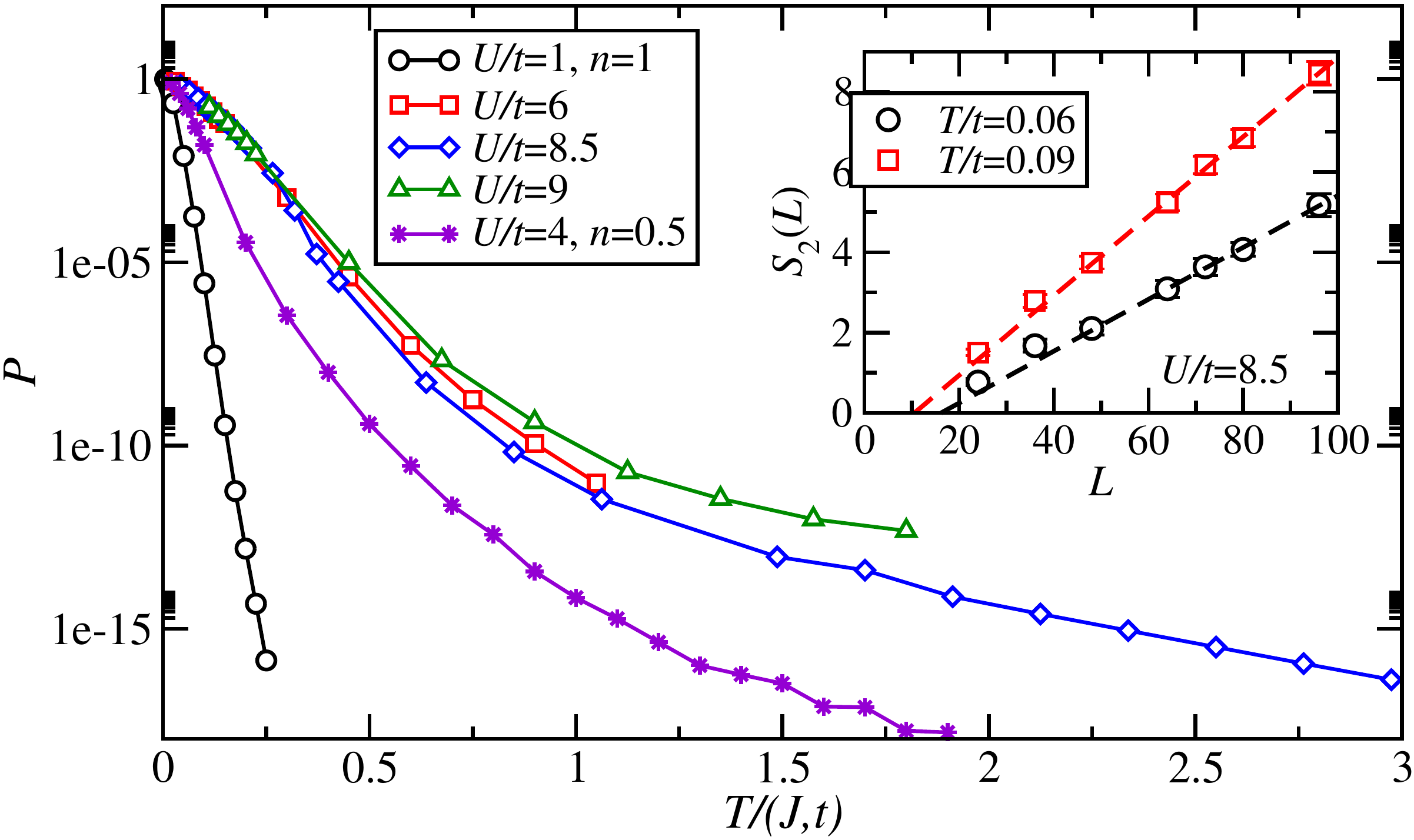}
\caption{(Color online) Purity $P$ as a function of $T/J$ ($T/t$) for the half-(quarter-) filled chain with $L=48$ sites.
The error bars are omitted for clarity of the figure.
\textit{Inset:} System size scaling of the purity for the Mott insulator with $U/t=8.5$. 
}
\label{fig:Purity}
\end{figure}

First, let us focus on the Mott insulating case at half filling.
Superexchange leads to an effective description in terms of a spin model and the ground state exhibits $2\kF$ correlations in the spin channel that are also reflected in the behavior of the Renyi entropy~\cite{laflorencie06}.
\Fref{fig:EntropyProfileU1} exemplifies the $n=2$ Renyi entropy profiles for $U/t=1$ (see \Fref{fig:EntropyScalingU6} for $U/t=6$).
For $T/t \lesssim 0.02$ the data for $L=48$ agrees perfectly with the ground state DMRG results and the system is effectively at zero temperature.
A second feature in the low-temperature crossover regime is the reduction of the amplitude of the $2\kF$ parity oscillation in the Renyi entropy.
We use a heuristic fitting ansatz where the thermal and quantum contribution to the entropy are simply added (including subleading oscillations).
This fit shows that the amplitude of the oscillations is suppressed exponentially in $T/t$ compatible with the presence of a finite thermal correlation length.
Above $T/t \gtrsim 0.2$, however, the oscillations are absent within our numerical accuracy but at the very boundary of the chain and the entropy increase is linear with blocks size.
The system is then in a regime dominated by thermal fluctuations although the temperatures for the data shown in \Fref{fig:EntropyProfileU1} are still below the superexchange scale.

Upon increasing the temperature, the entropy profile becomes asymmetric across the center of the system and the purity $P=\exp[- S_2(L)]$ starts to decrease.
It can be directly read of from the entropy profiles and we summarize its temperature dependence for different values of the interaction and $L=48$ in \Fref{fig:Purity}.
For larger values of $U/t$ we find that the purity for $T/J \lesssim 0.5$ is independent of the interaction but shows a common decrease that is exponential in $T/J$ above the finite-size gap.
This can be directly traced back to the effective spin physics in the low energy sector that depends on the spin velocity that is approximately proportional to $J$ for large $U$.
At $T \approx J$, $P$ begins to flatten and the systems enters the regime where the spin channel is disordered but the charge channel remains frozen since the temperature is still small compared to the interaction energy.
For $U/t=1$, however, things look differently because \textit{i)} the spin velocity is about a factor of three larger as compared to $U/t=9$ and \textit{ii)} the extent of the system is still too small compared to the charge correlation length such that the system in only on the verge of being a $c=1$ Mott insulator.
The system size dependence of the purity is exemplified in the inset of \Fref{fig:Purity} where the extensivity of the thermal entropy directly leads to an exponential suppression of $P$ as a function of $L$.
A detailed analysis of the connection between $S_2(L)$ and $v_s$ will be presented in \Sref{sec:velocities} below.

\begin{figure}[t]
\includegraphics[width=\columnwidth]{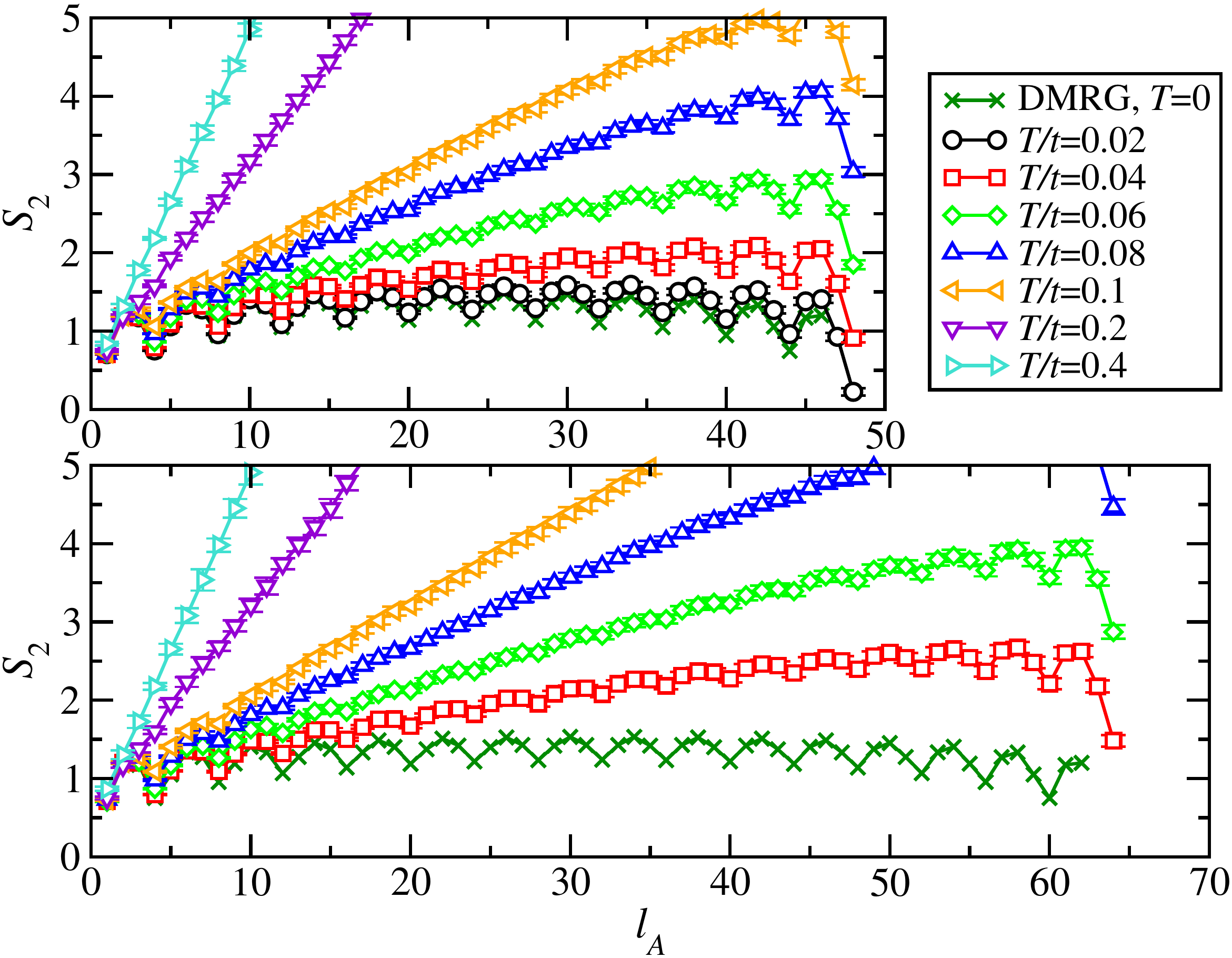}
\caption{(Color online) $n=2$ Renyi entropy profile for the quarter filled Hubbard chain at quarter filling and $U/t=4$.
The upper and lower panel correspond to 48 and 64 lattice sites respectively.
}
\label{fig:EntropyQuarterFilling}
\end{figure}

The $S_2$ profile for the metallic case is given in \Fref{fig:EntropyQuarterFilling} where we focus on the quarter-filled case.
The picture is similar compared to the Mott insulator regime.
For system sizes larger than 48, however, we do not reach the (effective) ground state even for $T/t=0.02$ opposed to the $U/t=1$ data shown in \Fref{fig:EntropyProfileU1}.

If the two velocities separate and, in particular, the charge velocity is much larger compared to $v_s$, it is possible to access the spin-incoherent Luttinger liquid regime~\cite{fiete07,feiguin10,feiguin11} where the spin channel is thermally disordered but the charge degrees of freedom can effectively be described by their low-energy field theory.
These effects will be reflected in a change of slope of the purity once the spin channel is completely disordered.

\subsection{Extracting $v_s$ and $v_c$}
\label{sec:velocities}
The  low-energy thermodynamics of a Luttinger liquid is governed by the spin and charge velocity and the prefactor of the volume law for the Renyi entropies encodes this information, as seen in \Eref{eq:ThermalRenyi}.
Since we do not know an exact expression for the Renyi profiles at finite temperature, we extract the velocities by either considering the entropy of the full system and the half system.
The latter quantity is most insensitive towards boundary effects and logarithmic contributions from the quantum entanglement.

\begin{table}[t]
 \begin{tabular}{|l|l|c|c|c|}
\hline
 $n$ & $U/t$	& QMC, $S(L/2)$ & QMC, $S(L)$ & exact~\cite{takahashi74,schulz94} \\ \hline
    1 &6	& $0.92 \pm 0.05$ & $0.937 \pm 0.01$ & 0.9257 \\ \hline
    1 &8.5      & $0.702 \pm 0.005$	& $0.720 \pm 0.006$  & 0.6929 \\ \hline
    1 &9	& $0.63 \pm 0.08$ & $ 0.64 \pm 0.02$  & 0.6588 \\ \hline \hline 
 1/2  &4        & $0.59 \pm 0.01$   & $0.6 \pm 0.1$ & 0.5961 \\ \hline
 \end{tabular}
\caption{$v_s$  and $1/[1/v_s+1/v_c]$ extracted from the Renyi entropies for the Mott insulator  and the quarter-filled metallic case from QMC simulations using either the half-system or the full-system entropy compared to the exact results.}
\label{tab:vs}
\end{table}

\begin{figure}[t]
\includegraphics[width=\columnwidth]{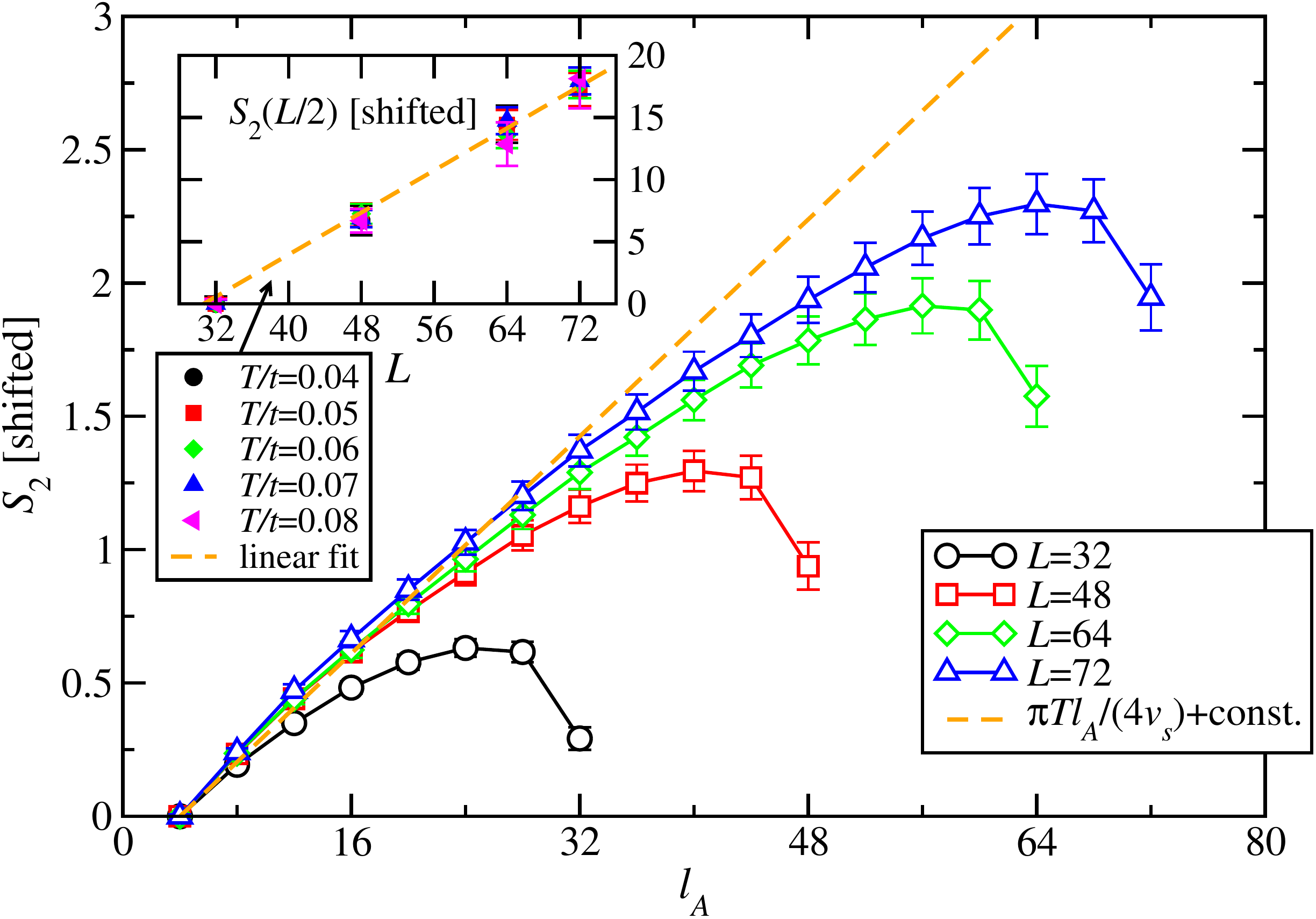}
\caption{(Color online) $S_2$ as a function of block size for the half-filled chain at $U/t=6$ and $T/t=0.06$ using a grid of $\delta l=4$.
The entropy data has been shifted such that $S(\delta l)=0$.
The orange dashed line is a guide for the eye with a slope $\pi T l_A/(4 v_s)$ (see \Eref{eq:ThermalRenyi}) that would correspond to purely thermodynamic entropy.
\textit{Inset:} $S_2/T$ for a half-system bipartition ($l_A=L/2$) for different temperatures from $T/t=0.04$ to $0.08$. The orange dashed line is a linear fit to the data at $T/t=0.06$ from which we obtain the spin velocity to be $v_s=0.92 \pm 0.05$ (see also Table \ref{tab:vs}).
}
\label{fig:EntropyScalingU6}
\end{figure}

\Fref{fig:EntropyScalingU6} shows the Renyi entropies for $n=1$, $U/t=6$ and $T/t=0.06$ for different system sizes.
One can see that $S_2$ for large block sizes eventually converges towards the thermal volume law as it is expected from \Eref{eq:vNHubbard} and (\ref{eq:ThermalRenyi}).
Fixing the lattice position $L/2$, $S_2/T$ is independent of temperature over a wide range of system sizes, as shown in the inset of \Fref{fig:EntropyScalingU6}.
Using a linear fit for system sizes $L>32$, we obtain the spin velocities from our ansatz in  \Eref{eq:ThermalRenyi} tabulated in Table \ref{tab:vs}.
A similar procedure is used to extract the velocities from the full system entropy (see inset in \Fref{fig:Purity}).
For the quarter filled system, we simply add the thermal contributions from the spin and charge sector.
In that way, we can not discriminate the spin and charge velocities separately but their inverse sum, $1/(1/v_s+1/v_c)$, also given in Table \ref{tab:vs}.
In fact, we find good agreement between analytic results and our QMC calculations justifying our fitting ansatz.

\subsection{Mutual Information}
\label{sec:resMI}
\begin{figure}[t]
\includegraphics[width=\columnwidth]{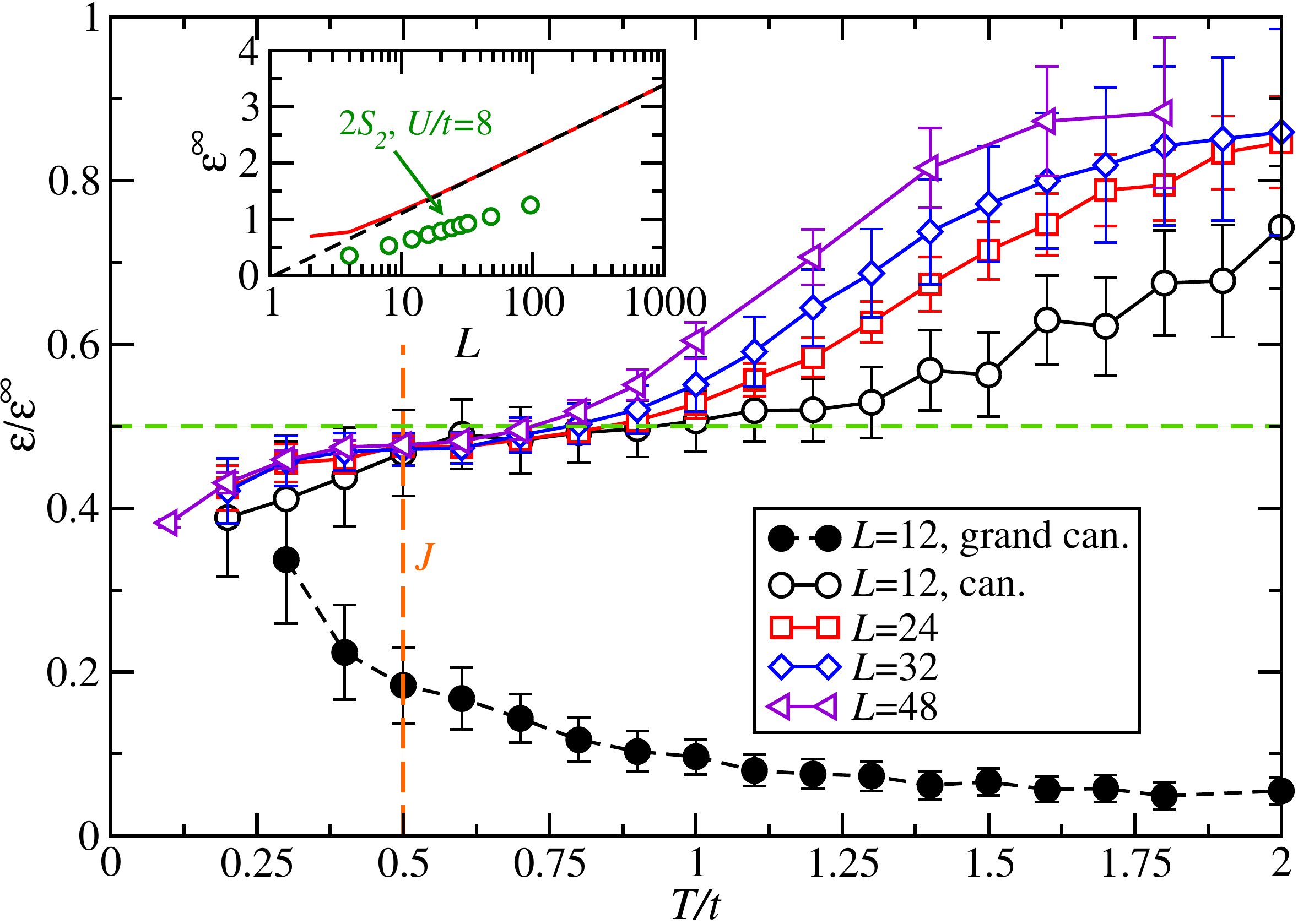}
\caption{(Color online) Mutual information for a bipartite splitting ($\Delta=0$) as a function of temperature for a Mott insulator at $U/t=8$.
The filled symbols correspond to data grand-canonical data with on average one particle per site.
The data is normalized by $\epsilon_\infty$ (see text) and the vertical dashed line marks the superexchange scale $J$.
\textit{Inset:}
$\epsilon^\infty$ for a half-filled system as a function of system size.
The black dashed line denotes the asymptotic scaling $\epsilon^\infty =1/2 \log L + \mathrm{const.}$ for large $L$.
The green data points correspond to $2S_2$ for the ground state at $U/t=8$.
}
\label{fig:excessEntropy}
\end{figure}

We calculate the mutual information $I_2$ between two equally large blocks pinned at the left and right boundary.
The blocks are grown simultaneously and have a size $L/2-\Delta/2$ where $\Delta$ is the separation of the blocks (see \Fref{fig:excessEntropy} for an illustration of the geometry).
$I_2$ for the quarter-filled chain at $U/t=4$ is shown in \Fref{fig:excessEntropy} for different system sizes and temperatures.
First of all, $I_2$ is sensitive towards the $2 \kF$ oscillations that disappear for larger temperatures as it was observed already in \Fref{fig:EntropyQuarterFilling} for the Renyi entropies.
As a function of block separation, the mutual information is suppressed exponentially at finite $T$ due to the presence of a finite correlation length and thus the correlations between two subsystems at, say $T/t=0.5$, only extend over a few lattice sites.

\begin{figure}[t]
\includegraphics[width=\columnwidth]{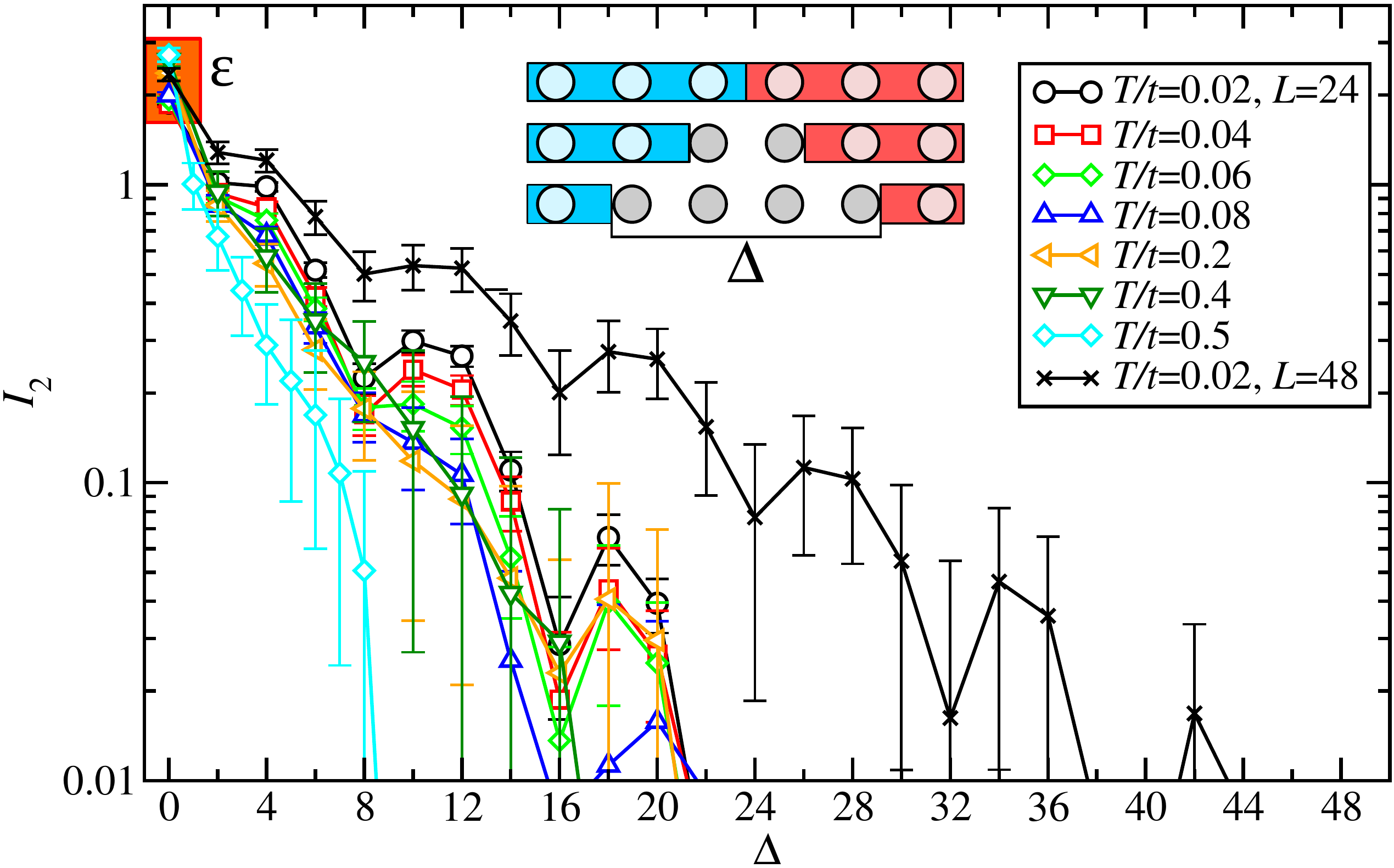}
\caption{(Color online) Mutual information $I_2$ for a quarter-filled Hubbard chain at $U/t=4$ and finite temperatures.
The inset illustrates the geometry of the two blocks that grow simultaneously with $\Delta$ sites between their edges, i.e. the block size is $L/2-\Delta/2$.
The orange shaded area denotes the mutual information for a bipartition, $\epsilon$, that is detailed for the Mott insulator in \Fref{fig:excessEntropy}.
}
\label{fig:MutInf}
\end{figure}

From \Fref{fig:MutInf} one can also see that the mutual information at $\Delta=0$, henceforth denoted as $\epsilon$, increases with temperature.
A more detailed analysis, shown for the Mott insulator in \Fref{fig:excessEntropy}, reveals an interesting structure.
Starting from the well known $T=0$ limit, $\epsilon$ increases up to temperatures on the order of the superexchange scale, $T\approx J$, where it exhibits a plateau.
For $T \gtrsim 2J$, $\epsilon$ shows a rapid increase again until it saturates eventually for temperatures much larger than $U$.
This behavior seems counterintuitive at first sight because we expect that thermal fluctuations destroy correlations between the subblocks since the correlation functions become more and more short ranged.
In particular, the infinitely hot state is maximally mixed and the (reduced) density matrix is proportional to the unit matrix, $\rho^\infty(A) = Z^{-1} \ident_Z$ where $Z=4^{l_A}$ is the partition function and $\ident_d$ is $d \times d$ unit matrix.
The systems considered here, however, are constrained by the total number of particles and zero net spin thus the particle number fluctuations of the two subblocks can not be totally uncorrelated.
The effect of particle number conservation on the complexity of infinitely hot states has been discussed in the context of matrix product operators~\cite{muth11,charrier12} where it is found that representing $\rho^\infty$ requires a finite matrix rank and the operator space entanglement entropy~\cite{prosen07} $S_\#$ of $\rho^\infty(A)$ diverges logarithmically, i.e. $S_\# = 1/2 \log L + \mathrm{const}$.
In fact, we see that $\epsilon$ for the unconstrained system -- in \Fref{fig:excessEntropy} we show data from grand-canonical simulations where the average particle number is one -- is monotonically decreasing and tends to zero very quickly.

The reduced density matrix for a subblock $A$ of the $T=\infty$ state, $\rho^\infty_A$, and henceforth the mutual information of the infinitely hot state denoted by $\epsilon^\infty$ in the presence of the particle and spin constraints can be obtained by simple combinatorial arguments.
For ease of simplicity, we focus on the case of a half system bipartition and unit filling but the results can be readily generalized to arbitrary fillings or partitions.
First, we realize that the density matrix is a direct product of the spin components thus it suffices to consider one species of filling $L/2$ and we restrict $\rho^\infty_A$ to one spin sector.
In the second step, we write the $\rho^\infty_A$ as a sum over the number of particles contained in block A,
$\rho^\infty_A = Z^{-1} \bigoplus_{n=0}^{L/2} \rho^{(n)}_A$, where each subblock is given as
\begin{equation}
 \rho^{(n)}_A = p_{L/2}(L/2-n) \ident_{p_{L/2}(n)}. 
\end{equation}
$p_m(n)=\binom{m}{n}$ is the number of possibilities to put $n$ particles onto $m$ lattice sites and $Z = p_{L}(L/2)$ is the partition function.
The $n$th moment can then be calculated as 
\begin{equation}
\tr [\rho^\infty_A]^\alpha = Z^{-\alpha} \sum_{n=0}^{L/2} [p_{L/2}(L/2-n) ]^\alpha p_{L/2}(n).
\end{equation}
The Renyi entropy and thus $\epsilon^\infty$ can then be evaluated using the definitions in Equations \ref{eq:Renyi} and \ref{eq:mutInf}.
The result is shown in the inset of \Fref{fig:excessEntropy} comparing it to the $T=0$ result for $U/t=8$.
In particular, it is straightforward to show that $\epsilon^\infty$ diverges like $(\alpha-1)/2 \log L$ for large $L$ where $\alpha$ is the Renyi index.

Turning back to the results in the main panel of \Fref{fig:excessEntropy} where we normalized $\epsilon$ with respect to $\epsilon^\infty$, we see that the plateau at $T \approx J$ corresponds to $\epsilon \approx \epsilon^\infty/2$.
This value is expected if the spin degrees of freedom are completely disordered but the charge channel is still gapped out.
Thus measuring $\epsilon/\epsilon^\infty$ in the canonical ensemble can provide us with direct information about the relative energy scales in the problem and whether we have some correlation apart from the constraints between the spin degrees of freedom.
For large temperatures, $U$ sets the energy scale for the saturation to $\epsilon/\epsilon^\infty=1$.

To probe this effects with cold atom experiments using the scheme based on a atomic gas microscope \cite{bakr09,sherson10} one should note that the scheme proposed there does not allow a post selection according to the particle number. If the total number of particles fluctuates from tube to tube this will automatically contribute to the measured entropy, as for example in the gran-canonical ensemble. Thus, to see effects that hinge on the constraint of fixed particle number,  such as the increase of the mutual information with increasing temperature discussed above, one has to ensure that the particle number fluctuations between pairs of tubes are small, e.g. as a consequence of the preparation process. For example one could prepare the state adiabatically, starting in a state with a pre-selected particle number. The latter can be achieved e.g. with techniques  realized in Ref.~\onlinecite{weitenberg11}.
This effect is also important in the context of particle number fluctuations, used for instance in measures like bipartite fluctuations~\cite{song12,rachel12}, where the constraint will also affect the finite-temperature behavior.

\section{Discussion}
We used recent QMC methods to access the finite-temperature crossover of the one-dimensional Hubbard model.
We can explore the domain up to very high temperatures where the entropy becomes extensive.
In particular, we accessed the second Renyi entropy to characterize quantum entanglement and thermodynamic dominated regimes by explicitly comparing to ground state DMRG data.
Characteristic features such as the $2\kF$ oscillations are suppressed rapidly
and the purity is exponentially small both in temperature as in system size.
Using the connection to the low-energy thermodynamics of the Luttinger liquid description, the spin and charge velocities can be extracted from the thermal contribution to the Renyi entropy.
In view of future experiments, it is thus possible to see features of entanglement in the Renyi profiles for not too large systems and extract vital information about the underlying Luttinger liquid using a measurement scheme that is solely based on single-site resolved density measurements~\cite{pichler13}.
An important issue, that is beyond the scope of this investigation, is an analytical description of the thermal crossover regime of the Renyi entropies for finite systems and we leave it for further studies.

We also showed the mutual information for different block sizes that is also very short-ranged for intermediate temperatures and shows similar features as the plain Renyi entropies.
In the presence of constraints, we also demonstrate that the bipartite mutual information can actually increase as a function of temperature and carries vital information about the nature of the fluctuations of the constrained and unconstrained system.

\section*{Acknowledgments}
We thank P. Zoller and A. J. Daley for useful discussions and acknowledge support by the Austrian Science Fund (FWF) through the SFB FoQuS (FWF Project No. F4018-N23)
and the Austrian Ministry of Science BMWF as part of the UniInfrastrukturprogramm of the Forschungsplattform Scientific Computing at LFU Innsbruck.


\end{document}